\documentclass[letter]{aa}  

\usepackage{graphicx}
\usepackage{txfonts}
\usepackage{lipsum}
\usepackage{subcaption}         
\usepackage{lscape}             
\usepackage{placeins}           
                                
\usepackage{natbib}
\bibpunct{(}{)}{;}{a}{}{,}
\providecommand{\sorthelp}[1]{}
\usepackage[colorlinks=true, allcolors=blue]{hyperref}

\usepackage{xcolor}

\begin{document}


   \title{Discriminating {\it Planck} reionisation histories with the kSZ effect}


   \titlerunning{Discriminating  {\it Planck} reionisation histories with the kSZ effect}

  \author{  M. Douspis\inst{1,2} \thanks{marian.douspis@universite-paris-saclay.fr}
     \and
     A. Gorce\inst{1}
     \and
    S. Ilić\inst{3}
    \and 
    L. McBride\inst{4}
     \and
     M. Muñoz-Echeverría\inst{6}
    \and
     E. Pointecouteau\inst{7}
     \and
     L. Salvati\inst{1}
     \and
     M. Tristram\inst{5}
     }

  \institute{Université Paris-Saclay, CNRS, Institut d’Astrophysique Spatiale, 91405 Orsay, France
    \and
    Observatoire de Paris, PSL Research University, Sorbonne Université, CNRS, LUX, 75014 Paris, France
    \and
    Institut du Développement et des Ressources en Informatique Scientifique (IDRIS), CNRS, Université Paris-Saclay, Orsay, France
    \and
    Université Paris-Saclay, Inria, Inria Saclay-Île-de-France, 91120, Palaiseau, France.
    \and 
    IJCLab, Université Paris-Saclay, CNRS/IN2P3, IJCLab, 91405 Orsay, France
    \and 
    Institut de Radioastronomie Millimétrique (IRAM), Avenida Divina Pastora 7, Local 20, E-18012, Granada, Spain
    \and
    IRAP, CNRS, Université de Toulouse, CNES, UT3-UPS, Toulouse, France
    }

   \date{Received December 25, 20XX}

  \abstract{
The epoch of reionisation is a key phase in cosmic history, but its detailed evolution remains poorly constrained by current cosmic microwave background (CMB) observations. We investigated whether the kinetic Sunyaev--Zel’dovich (kSZ) effect can discriminate among reionisation histories consistent with current large--scale CMB constraints. Using histories derived from \textit{Planck} data, we computed the corresponding kSZ angular power spectra within an analytical framework. 
The allowed histories fall into two broad classes, `late' and `early' start, yielding distinct kSZ signatures that remain clearly separable even when accounting for modelling uncertainties in both the reionisation scenario, $x_e(z)$, and the properties of early galaxies. Current kSZ measurements ($\sim$0--3 $\mu$K$^2$) tend to favour `late' reionisation models but are not yet sensitive enough to definitively distinguish between the scenarios -- a measurement of the kSZ power spectrum at $\ell \sim 2000$ with $\sim$0.4 $\mu$K$^2$ sensitivity, achievable in the coming years, will be sufficient to do so. This work demonstrates that CMB data alone can constrain the reionisation midpoint $z_\mathrm{re}$ with extremely narrow error bars ($7.94<z_\mathrm{re}<8.17$), even when effectively marginalising over modelling uncertainties.

}

   \keywords{Reionisation --
                CMB --
                kinetic SZ
               }

   \maketitle
   \nolinenumbers

\section{Introduction}

The epoch of reionisation (EoR) marks a pivotal phase in cosmic history, during which the first galaxies emitted sufficient ultraviolet radiation to ionise the intergalactic medium (IGM), ending the cosmic Dark Ages. Understanding the timing, duration, and sources of reionisation remains a key challenge of modern cosmology. Observations of the cosmic microwave background (CMB) by the {\it Planck} satellite, in temperature and polarisation, have provided crucial constraints on the optical depth to reionisation \citep{planck2014-a25,Ilic25}, offering an overall view of this process. However, these measurements are limited by the integrated nature of the signal, thus only allowing us to measure a range of possible reionisation histories.

Recent observations with high-resolution CMB experiments have opened a new window on the EoR through the measurement of the kinematic Sunyaev-Zel’dovich effect \citep[kSZ; ][]{kSZ1, kSZ2} at small angular scales. The kSZ effect, arising from the scattering of CMB photons by free electrons in moving ionised gas, is sensitive to the distribution and velocity of ionised matter during and after the EoR. By probing the kSZ signal at arcminute scales, upcoming and current CMB experiments can, in theory, constrain the duration, patchiness, and inhomogeneity of reionisation, complementing {\it Planck}’s large-scale data \citep[e.g.][]{Zahn2012_reio, planck2014-a25, kSZmeasurement2, ACTBeringue, Chaubal26}. 

In this work, we investigate whether the kSZ signatures of the reionisation histories allowed by the latest \textit{Planck} data \citep{Ilic25} are distinct enough for high-resolution experiments to differentiate them. We assume a fixed cosmology throughout \citep[][`Planck (baseline)', Table~1]{Tristram25}.

 \begin{figure*}[!th]
   \sidecaption
   \includegraphics[width=0.6\hsize]{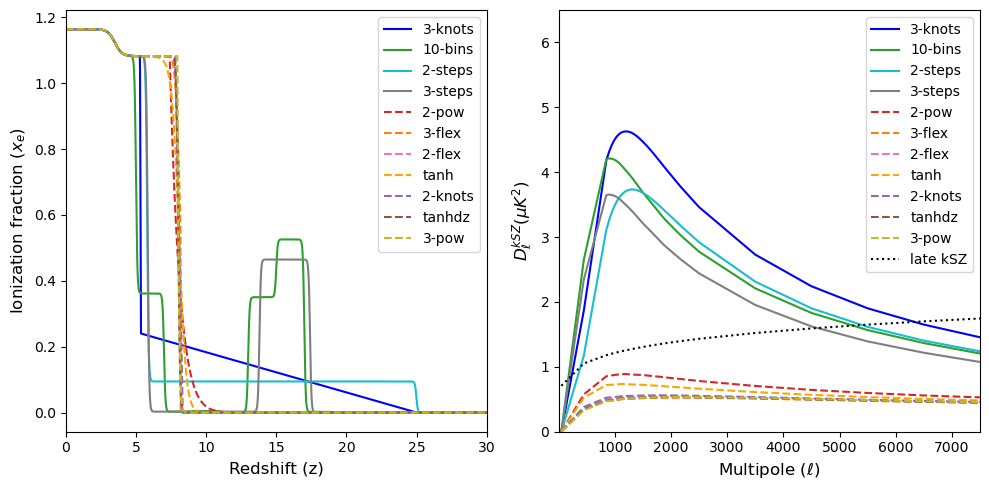}
      \caption{Reionisation history (left) and corresponding patchy- (solid and dashed) and late- (dotted) kSZ angular power spectra allowed by \textit{Planck} PR4 \citep{Ilic25}. Solid (dashed) lines are used for the `early' (`late') reionisation histories. Labels are defined in the text. }
         \label{Fig:fig1}
  \end{figure*}


\section{The kSZ angular power spectrum}

The kSZ effect has two distinct contributions. One is a late-time component arising from bulk flows of ionised gas after the end of reionisation. The other is a patchy component encoding information about the timing, duration, and morphology of cosmic reionisation that is generated as ionised bubbles form and expand in the IGM. 

\subsection{Modelling patchy kSZ\label{sec:model}}

To model the patchy kSZ angular power spectrum (APS), we followed the analytical approach of \cite{Gorce20}. It relies on two main ingredients, within the Limber approximation: the reionisation history, $x_e(z)$, and the electron overdensity power spectrum, whose shape is parametrised by an amplitude at large scales, $\alpha_0$, and a cut-off scale ($\kappa$) loosely related to a typical ionised bubble size.

\subsubsection{Histories of reionisation from {\it Planck}\label{sec:xez}}

\cite{Ilic25} used the latest {\it Planck} PR4 CMB data \citep{planck2020-LVII} to constrain the reionisation history of the Universe, systematically evaluating a wide range of models, from simple parametric to flexible non-parametric approaches, with 1 to 12 free parameters. Their analysis, which employed both Bayesian and frequentist approaches, yielded robust and consistent estimates of the reionisation optical depth. Using the frequentist approach, the authors obtained $\tau = 0.0581 \pm 0.0061 \pm 0.0005$, where the first uncertainty represents the statistical error and the second corresponds to the systematic error associated with $x_e$ modelling. This systematic uncertainty, together with that arising from the assumed foreground models \citep{Tristram25}, is negligible in comparison with the current statistical uncertainty.
These results suggest that reionisation is a relatively recent process (see Fig.~\ref{Fig:fig1}) -- a conclusion reached in other studies based on CMB data \citep[e.g.][]{planck2014-a25,planck2016-l06,2020A&A...635A..99P, 2023A&A...675A..12P,planck2020-LVII,2026arXiv260322454G}. We chose to consider the best-fit $x_e(z)^\mathrm{best}$ for each model obtained with the frequentist approach in \cite{Ilic25} as our inputs when deriving the kSZ APS.\footnote{We excluded the 
\texttt{pca-5} parameterisation from \cite{Ilic25}, as it contains negative values, and thus we were left with 11 models.}.

We labelled the parameterisations following \cite{Ilic25}: hyperbolic tangent models with one or two free parameters (\texttt{tanh}, \texttt{tanhdz}); power-law models with two or three free parameters (\texttt{pow-2}, \texttt{pow-3}); step-based models with two or three steps (\texttt{2-steps} and \texttt{3-steps}); interpolation-based models with two or three nodes (\texttt{2-knots}, \texttt{3-knots} for the linear interpolation, and \texttt{2-flex}, \texttt{3-flex} for the PCHIP interpolation); and binned models (\texttt{10-bins)}. 
The corresponding $x_e(z)$ histories are shown in Fig.~\ref{Fig:fig1} (left) and can be divided into two groups according to how early or late they first reach a 10\% ionisation fraction, as quantified by $z_{10}^{first}$, which is the redshift at which this threshold is first attained. Models with early (late) reionisation are defined with $ z_{10}^{first} > 15 $ ($z_{10}^{first} < 10 $). Both groups can be characterised further in terms of the contribution of high redshifts ($z>10$) to the overall optical depth. In early models, about one fourth of $\tau$ comes from such redshifts (see Table ~\ref{table:A1} ).

\paragraph{Modelling uncertainties.}

To propagate the uncertainty on the history of reionisation to the kSZ APS, for each parameterisation we used all $x_e(z)$ models that satisfy $\chi^2 < \chi^2(x_e(z)^\mathrm{best})+1$ \citep[dashed red line intervals in Fig.~6 of][]{Ilic25}. The resulting ensemble envelop on the derived kSZ APS is defined as the 68\% confidence interval around the fiducial kSZ spectrum.

\subsubsection{Electron overdensity for LORELI II}

The parameters $\alpha_0$ and $\kappa$ required to derive the kSZ APS are still unconstrained by data, so we chose to consider the range of possibilities provided by the LORELI II suite of N-body hydro-radiative simulations \citep{MeriotSemelin_2024,MeriotSemelin_2025}. The $\sim$10\,000 simulations were generated by varying five astrophysical parameters that impact the ionising properties of galaxies.  Two parameters that describe their X-ray production have been shown to have little impact on the kSZ spectrum amplitude and shape \citep[see Appendix C in ][]{McBride25} and were thus fixed. The three remaining parameters describe the source properties: the gas conversion timescale governs how efficiently galaxies turn gas into stars, the minimum halo mass defines the critical mass above which the star formation is sufficient to produce enough ionising radiation to form bubbles, and the ionising escape fraction is defined by the proportion of photons not absorbed and thus escaping to reionise the IGM. 
By spanning a wide range of these parameters, the LORELI II simulations are representative of a broad range of plausible reionisation scenarios with various reionisation histories and morphologies, resulting in distinct kSZ spectra.\footnote{We assumed that reionisation histories and ($\alpha_0,\kappa$) are uncorrelated.}

From the 1106 simulations in which the gas conversion timescale, the minimum halo mass, and the ionising escape fraction were varied, we fit for the overall amplitude ($\alpha_0$) and cut-off scale ($\kappa$) on the electron overdensity power spectra following the procedure described in \citet{Gorce20}. 
We further selected the set of pairs ($\alpha_0, \kappa$) corresponding to LORELI II simulations falling within the 1$\sigma$  confidence interval of the optical depth $\tau$ defined in Sect.~\ref{sec:xez}, leaving us with 319 simulations. For our fiducial models, we used the averages over this subset: $\hat \alpha=\langle \log \alpha_0/\mathrm{Mpc^3} \rangle =4.8$ and $\hat \kappa = \langle \kappa \rangle =0.03\;\mathrm{Mpc^{-1}}$. 

We computed 11 fiducial kSZ angular power spectra by assuming (i) the frequentist best-fit models of \cite{Ilic25}, corresponding to 11 parameterisations of $x_e$ fitted to {\it Planck} PR4 data, and (ii) the mean values ($\hat\alpha$, $\hat\kappa$) for all models. The resulting spectra are shown in Fig.~\ref{Fig:fig1} (right).

\paragraph{Modelling uncertainties.\label{sec:astro}}

The 319 $(\alpha_0, ~\kappa)$ pairs define the range of possible kSZ APS arising from uncertainties in the astrophysical properties of the first sources, namely the dispersion of $\alpha_0$ and  $\kappa$ in the LORELII~II simulations. For a given history of reionisation, we computed all the 319 kSZ APS. The resulting 68\% confidence envelop represents our modelling uncertainties.

\subsection{Late-time kSZ}

For the late-time angular power spectrum component, we assumed the analytical shape of \cite{Park_Alvarez_Bond_2018}, which is in perfect agreement with the simulation of \cite{Shaw12}. We rescaled the amplitude following the scaling relation of \cite{Shaw12}, assuming $\tau = 0.058$, $\sigma_8=0.8073$, and $h=0.6777$. 
The best-fit values of $\tau$, $\sigma_8$, and $H_0$ vary slightly across the $x_e(z)$ models considered \citep[see Fig.~5 in][]{Ilic25}. However, we found the resulting variations in the late-time kSZ APS to be negligible ($<4\%$) compared to variations in the patchy spectra. Our late time kSZ APS is shown as a dotted line in Fig.~\ref{Fig:fig1}.

\section{kSZ signatures of {\it Planck} histories of reionisation
}

 \begin{figure*}[!ht]
   \centering
   \includegraphics[width=0.96\hsize]{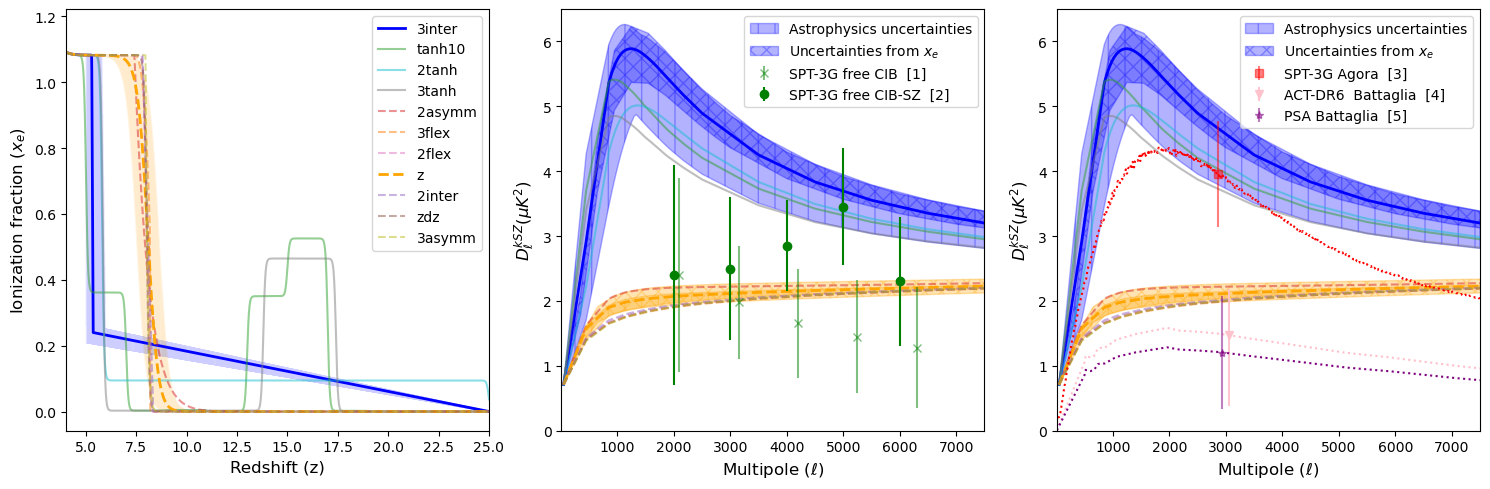}
      \caption{Same as Fig.~\ref{Fig:fig1} but with the two selected reference histories (\texttt{tanh} and \texttt{3-knots}) representing the ‘late’ and ‘early’ groups highlighted in orange and blue, respectively. \textit{Left:} Reionisation histories. Uncertainties from the {\it Planck} analysis are shown as shaded regions. \textit{Middle and right:} Total (patchy + late) kSZ power spectra, with uncertainties propagated from the $x_e$ uncertainties shown in the left panel (cross-hatched regions), and additional uncertainties arising from the LORELI II–allowed ranges of $\alpha_0$ and $\kappa$ (vertical hatched regions). \textit{Middle:} kSZ spectra compared with binned estimates from SPT-3G. Right: kSZ spectra compared with fitted templates in ACT-DR6, SPT-3G and combined CMB analysis (see text for legends).}
         \label{Fig:fig2}
   \end{figure*}
   
\subsection{Fiducial kSZ}

We show in Fig.~\ref{Fig:fig1}  the 11 reionisation histories $x_e(z)$ corresponding to the best-fit models of \cite{Ilic25} (left) as well as the corresponding kSZ angular power spectra, with their patchy (solid and dashed lines) and late (dotted) components, computed with the fiducial $\hat{\alpha}$ and $\hat{\kappa}$ values (right). The kSZ power spectra exhibit two behaviours depending on the duration of the reionisation \citep[as expected: see e.g.][]{Shaw12, Gorce20}. We highlight the difference between these two sets in Fig.~\ref{Fig:fig1} by showing in dashed lines the `late' reionisation models and in solid lines the `early' reionisation ones. The total (late + patchy) kSZ APS shown in Fig.~\ref{Fig:fig2} exhibit the same behaviour as the patchy kSZ, with early (late) history spectra dominated by the patchy (late-time) contribution.

This result indicates that a kSZ measurement could discriminate between these two broad categories. However, it was obtained for fixed astrophysical parameters.
Thus, in the following, we investigate whether the discriminating power of a kSZ measurement is robust to uncertainties on the reionisation history $x_e(z)$ and on the astrophysical parameters ($\alpha_0$, $\kappa$). To this end, we selected one reference model in each group: the hyperbolic tangent \texttt{tanh} and the three-point interpolation \texttt{3-knots}.

\subsection{Propagating $x_e$ uncertainties}

We propagated the uncertainties in $x_e$ defined in Sec.~\ref{sec:xez} by computing the kSZ angular power spectrum for all models within the 68\% interval and evaluating the dispersion at each multipole, as shown by the cross-hatched regions in Fig.~\ref{Fig:fig2} (middle and right).
The $\pm 0.006$ uncertainty in $\tau$ \citep{Ilic25} from the {\it Planck} data produces an uncertainty envelope for each kSZ spectrum, but the two groups are still distinguishable.

\subsection{Propagating astrophysical uncertainties\label{sec:res:astro}}

Our fiducial assumption for the shape of the electron overdensity power spectrum follows the work of \cite{Gorce20} and was derived from the mean values of ($\alpha_0$, $\kappa$) in our LORELI II subset (see Sec.~\ref{sec:astro}). However, this particular ($\alpha_0$, $\kappa$) set may not be representative of our Universe.
Thus, we computed, for our two reference early and late reionisation histories, the kSZ power spectra of the 319 ($\alpha_0$, $\kappa$) fitted to the LORELI II subset, and we show their dispersion in Fig.~\ref{Fig:fig2}. For both \texttt{tanh} and \texttt{3-knots}, we show the resulting uncertainty envelope at each multipole $\ell$ (vertical hatch in Fig.~\ref{Fig:fig2}).

We find that even after marginalising over the range of astrophysical models allowed by the LORELI II simulations, the two groups of reionisation histories remain clearly distinguishable. The resulting envelope widths are comparable to those obtained propagating $x_e$ uncertainties: Both sources of uncertainty contribute at a similar level.\footnote{Note that combining them would require elaborate propagation, as they are weakly degenerated} 

Measurements of the kSZ power spectrum could therefore provide a powerful discriminant among the reionisation histories allowed by {\it Planck}. 
However, our results are somewhat conservative and limited by the fact that we considered the two sources of uncertainties independently, while in reality reionisation history and galaxy properties are correlated. 
More robust estimates, for example marginalising over cosmological parameters or jointly accounting for the correlated uncertainties, are required to fully characterise the two-scenario behaviour.

\subsection{Potential discrimination by current data}

We now confront our findings with recent measurements of the kSZ angular power spectrum from the South Pole Telescope \citep[SPT;][]{Chaubal26}, the Atacama Cosmology Telescope \citep[ACT;][]{ACTLouis}, and a joint analysis of {\it Planck}, ACT, and SPT data \citep{Tristram25, ACTBeringue}.

The spectra obtained from fitting the amplitude of a kSZ template to data were compared to our envelopes of allowed spectra in the right panel of Fig.~\ref{Fig:fig2}. 
\cite{Chaubal26} analysed 90, 150, and 220~GHz observations over 1646~deg$^2$ from the SPT-3G survey using several templates of the kSZ signal and other extragalactic foregrounds. With the Agora template \cite[][in red]{Agora}, the data shows better agreement with the `early' reionisation models. In contrast, with the \citet{Battaglia} template, both the ACT-DR6 data alone \citep[][in pink]{ACTLouis} or the combination of \textit{Planck}, SPT, and ACT data \citep[][in purple]{Tristram25} favour a kSZ amplitude at $\ell = 3000$, which remains below $2.1\,\mu\mathrm{K}^2$ at 1$\sigma$,
even when marginalising on other foregrounds \citep[see][]{ACTBeringue,Tristram25}. 
The kSZ amplitude was also measured in $\ell$ bins in \citet{Chaubal26}, 
and the authors assumed a free shape for the cosmic infrared background (CIB) alone (green crosses) or for both the CIB and the kSZ\footnote{Labelled as free CIB $\ell^\alpha$ G15 SZ and free CIB+SZ in \cite{Chaubal26}, respectively.} (green circles). Both approaches yielded spectra that are fully consistent with `late' reionisation scenarios (Fig.~\ref{Fig:fig2}, middle panel).

Going beyond our reference \texttt{tanh} and \texttt{3-knots} $x_e$ models and assuming, for example, the five free CIB points measured by \citet{Chaubal26}, all late (early) reionisation histories lead to $\chi^2 < 2.5$ ($\chi^2 > 16$), as shown in Table~\ref{table:A1}. The same preference remains for all sets of measurements shown (except the one coming from the SPT data fit with the Agora template). Overall, most recent data favour a low kSZ power, which corresponds to `late' reionisation scenarios, although the dispersion between analyses and the associated uncertainties remain significant.

\section{Discussion and conclusions}

While \textit{Planck} provides robust constraints on the optical depth to reionisation, the detailed shape of the reionisation history remains uncertain. The set of allowed histories can be broadly divided into early and late scenarios, which lead to distinct kSZ power spectra.

Assuming a fixed cosmology corresponding to the concordance 
$\Lambda$CDM model, we investigated how robust this split is to modelling uncertainties propagated from (i) the determination of the best-fit reionisation history, $x_e(z)$, from large-scale CMB data and (ii) galaxy properties from the LORELI II simulation suite. Within 68\% confidence intervals, we find that these two uncertainty sources contribute comparably to the dispersion of the kSZ angular power spectra (about 10\%). However, the two classes of reionisation histories remain clearly distinguishable.

 Despite spanning a wide range of amplitudes (about $0-3~\mu$K$^2$), current measurements of the kSZ angular power spectrum are sufficient to discriminate between the two groups. Most current data favour late reionisation scenarios over early ones (Table~\ref{table:A1}), corresponding to $7.94<z_\mathrm{re}<8.17$ and $0.22<\Delta z<1.47$ constraints, even when effectively marginalising over modelling uncertainties. The two groups could eventually be discriminated at the 5$\sigma$ level with a future measurement of the total kSZ angular power spectrum at $\ell \sim 2000$ with a sensitivity of $\sim 0.4 \,\mathrm{\mu K^2}$.
 
 In order for kSZ measurements to further discriminate between individual reionisation histories within a subgroup and provide the detailed shape of the reionisation history as a function of redshift, observational data should reach at least a 10\% sensitivity to match variations between models within the same group.
A comprehensive analysis simultaneously probing cosmology, reionisation history, kSZ amplitude, and associated systematics would be required to reach such accuracy \citep[following for example][]{Douspis_2022_I,GorceDouspis_2022}.  \cite{Tristram25}  have recently developed a joint likelihood combining \textit{Planck} and ACT together with the latest SPT-3G measurements, which could be exploited to perform such an analysis and extract a maximum of information from current data before stage~4 and 5 telescopes go online.

Other observations, such as of the quasar damping wing, the Lyman-$\alpha$ forest, or high-redshift galaxy luminosity functions, may also soon help in discriminating between these scenarios. However, current constraints remain limited by degeneracies with poorly constrained astrophysical parameters \citep[e.g. $f_{\rm esc}$,][]{Whitler}. By distinguishing between reionisation histories, kSZ measurements could help break these degeneracies, while the high-redshift 21\,cm signal would ultimately reveal the full reionisation history.

\begin{acknowledgements}
      The authors thank R. Mériot and B. Semelin for providing the LORELI simulations and acknowledge the support of the French Agence Nationale de la Recherche (ANR), under grant ANR-22-CE31-0010 (project BATMAN). This work was supported by the ``action thématique'' Cosmology-Galaxies (ATCG) of the CNRS/INSU PN Astro and ANR PIA funding ANR-20-IDEES-0002. We also acknowledge the use of  open-source libraries, including \texttt{CAMB}~\citep{Lewis:1999bs},
      and \texttt{matplotlib} \citep{matplotlib}.
\end{acknowledgements}

\bibliographystyle{aa}
\bibliography{aa60915-26}

@ARTICLE{Ilic25,
       author = {{Ili{\'c}}, S. and {Tristram}, M. and {Douspis}, M. and {Gorce}, A. and {Henrot-Versill{\'e}}, S. and {Hergt}, L.~T. and {Langer}, M. and {McBride}, L. and {Mu{\~n}oz-Echeverr{\'\i}a}, M. and {Pointecouteau}, E. and {Salvati}, L.},
        title = "{Reconstructing the epoch of reionisation with Planck PR4}",
      journal = {\aap},
         year = 2025,
        month = aug,
       volume = {700},
          eid = {A26},
        pages = {A26},
          doi = {10.1051/0004-6361/202555196},
archivePrefix = {arXiv},
       eprint = {2504.13254},
 primaryClass = {astro-ph.CO},
       adsurl = {https://ui.adsabs.harvard.edu/abs/2025A&A...700A..26I}
}

@ARTICLE{Gorce20,
       author = {{Gorce}, A. and {Ili{\'c}}, S. and {Douspis}, M. and {Aubert}, D. and {Langer}, M.},
        title = "{Improved constraints on reionisation from CMB observations: A parameterisation of the kSZ effect}",
      journal = {\aap},
         year = 2020,
        month = aug,
       volume = {640},
          eid = {A90},
        pages = {A90},
          doi = {10.1051/0004-6361/202038170},
archivePrefix = {arXiv},
       eprint = {2004.06616},
 primaryClass = {astro-ph.CO},
       adsurl = {https://ui.adsabs.harvard.edu/abs/2020A&A...640A..90G}
}

@ARTICLE{McBride25,
       author = {{McBride}, Lisa and {Gorce}, Ad{\'e}lie and {Douspis}, Marian and {Meriot}, Romain and {Semelin}, Beno{\^\i}t and {Hergt}, Lukas T. and {Ili{\'c}}, Stephane and {Mu{\~n}oz-Echeverr{\'\i}a}, Miren and {Pointecouteau}, Etienne and {Salvati}, Laura and {Tristram}, Matthieu},
        title = "{Astrophysical constraints from future measurements of the kinetic Sunyaev-Zel'dovich power spectrum}",
      journal = {arXiv e-prints},
         year = 2025,
        month = nov,
          eid = {arXiv:2511.22309},
        pages = {arXiv:2511.22309},
          doi = {10.48550/arXiv.2511.22309},
archivePrefix = {arXiv},
       eprint = {2511.22309},
 primaryClass = {astro-ph.CO},
       adsurl = {https://ui.adsabs.harvard.edu/abs/2025arXiv251122309M}
}

@ARTICLE{Shaw12,
       author = {{Shaw}, Laurie D. and {Rudd}, Douglas H. and {Nagai}, Daisuke},
        title = "{Deconstructing the Kinetic SZ Power Spectrum}",
      journal = {\apj},
         year = 2012,
        month = sep,
       volume = {756},
       number = {1},
          eid = {15},
        pages = {15},
          doi = {10.1088/0004-637X/756/1/15},
archivePrefix = {arXiv},
       eprint = {1109.0553},
 primaryClass = {astro-ph.CO},
       adsurl = {https://ui.adsabs.harvard.edu/abs/2012ApJ...756...15S}
}

@ARTICLE{MeriotSemelin_2024,
       author = {{Meriot}, R. and {Semelin}, B.},
        title = "{The LORELI database: 21 cm signal inference with 3D radiative hydrodynamics simulations}",
      journal = {\aap},
         year = 2024,
        month = mar,
       volume = {683},
          eid = {A24},
        pages = {A24},
          doi = {10.1051/0004-6361/202347591},
       adsurl = {https://ui.adsabs.harvard.edu/abs/2024A&A...683A..24M}
}

@article{MeriotSemelin_2025,
  title={Comparison of Bayesian inference methods using the LORELI II database of hydro-radiative simulations of the 21-cm signal},
  author={Meriot, Romain and Semelin, Benoit and Cornu, David},
  journal={\aap},
  volume={698},
  pages={A80},
  year={2025},
  publisher={EDP Sciences}
}

@ARTICLE{Chaubal26,
       author = {{Chaubal}, P. and {Huang}, N. and {Reichardt}, C.~L. and {Anderson}, A.~J. and {Ansarinejad}, B. and {Archipley}, M. and {Balkenhol}, L. and {Barron}, D.~R. and {Benabed}, K. and {Bender}, A.~N. and {Benson}, B.~A. and {Bianchini}, F. and {Bleem}, L.~E. and {Bocquet}, S. and {Bouchet}, F.~R. and {Bryant}, L. and {Camphuis}, E. and {Campitiello}, M.~G. and {Carlstrom}, J.~E. and {Carron}, J. and {Chang}, C.~L. and {Chichura}, P.~M. and {Chokshi}, A. and {Chou}, T.-L. and {Coerver}, A. and {Crawford}, T.~M. and {Daley}, C. and {de Haan}, T. and {Dibert}, K.~R. and {Dobbs}, M.~A. and {Doohan}, M. and {Doussot}, A. and {Dutcher}, D. and {Everett}, W. and {Feng}, C. and {Ferguson}, K.~R. and {Ferree}, N.~C. and {Fichman}, K. and {Foster}, A. and {Galli}, S. and {Gambrel}, A.~E. and {Gao}, A.~K. and {Gardner}, R.~W. and {Ge}, F. and {Goeckner-Wald}, N. and {Gualtieri}, R. and {Guidi}, F. and {Guns}, S. and {Halverson}, N.~W. and {Hivon}, E. and {Ho}, A.~Y.~Q. and {Holder}, G.~P. and {Holzapfel}, W.~L. and {Hood}, J.~C. and {Hryciuk}, A. and {Jhaveri}, T. and {K{\'e}ruzor{\'e}}, F. and {Khalife}, A.~R. and {Knox}, L. and {Korman}, M. and {Kornoelje}, K. and {Kuo}, C.-L. and {Levy}, K. and {Li}, Y. and {Lowitz}, A.~E. and {Lu}, C. and {Lynch}, G.~P. and {Maccarone}, T.~J. and {Maniyar}, A.~S. and {Martsen}, E.~S. and {Menanteau}, F. and {Millea}, M. and {Montgomery}, J. and {Nakato}, Y. and {Natoli}, T. and {Noble}, G.~I. and {Omori}, Y. and {Ouellette}, A. and {Pan}, Z. and {Paschos}, P. and {Phadke}, K.~A. and {Pollak}, A.~W. and {Prabhu}, K. and {Quan}, W. and {Raghunathan}, S. and {Rahimi}, M. and {Rahlin}, A. and {Rouble}, M. and {Ruhl}, J.~E. and {Schiappucci}, E. and {Silva Oliveira}, A.~C. and {Simpson}, A. and {Sobrin}, J.~A. and {Stark}, A.~A. and {Stephen}, J. and {Tandoi}, C. and {Thorne}, B. and {Trendafilova}, C. and {Umilta}, C. and {Vieira}, J.~D. and {Vieregg}, A.~G. and {Vitrier}, A. and {Wan}, Y. and {Whitehorn}, N. and {Wu}, W.~L.~K. and {Young}, M.~R. and {Zebrowski}, J.~A.},
        title = "{SPT-3G D1: A Measurement of Secondary Cosmic Microwave Background Anisotropy Power}",
      journal = {arXiv e-prints},
         year = 2026,
        month = jan,
          eid = {arXiv:2601.20551},
        pages = {arXiv:2601.20551},
          doi = {10.48550/arXiv.2601.20551},
archivePrefix = {arXiv},
       eprint = {2601.20551},
 primaryClass = {astro-ph.CO},
       adsurl = {https://ui.adsabs.harvard.edu/abs/2026arXiv260120551C}
}

@ARTICLE{Tristram25,
       author = {{Tristram}, M. and {Douspis}, M. and {Gorce}, A. and {Henrot-Versill{\'e}}, S. and {Hergt}, L.~T. and {Ilic}, S. and {McBride}, L. and {Mu{\~n}oz-Echeverr{\'\i}a}, M. and {Pointecouteau}, E. and {Salvati}, L.},
        title = "{Combining cosmic microwave background datasets with consistent foreground modelling}",
      journal = {\aap},
         year = 2026,
        month = jun,
       volume = {710},
          eid = {A165},
        pages = {A165},
          doi = {10.1051/0004-6361/202558015},
archivePrefix = {arXiv},
       eprint = {2511.04733},
 primaryClass = {astro-ph.CO},
       adsurl = {https://ui.adsabs.harvard.edu/abs/2026A&A...710A.165T}
}

@ARTICLE{planck2014-a25,
author = {{\sorthelp{Planck Collaboration IntZV}}{Planck Collaboration Int.
 XLVII}},
title = "{\textit{Planck} intermediate results. XLVII. Constraints on
 reionization history}",
journal = {\aap},
archivePrefix = "arXiv",
eprint = {1605.03507},
year = 2016,
volume = 596,
pages = {A108},
doi = {10.1051/0004-6361/201628897}
}

@ARTICLE{2026arXiv260322454G,
       author = {{Genesini}, Valentina and {Galloni}, Giacomo and {Pagano}, Luca and {Campeti}, Paolo and {Lattanzi}, Massimiliano},
        title = "{Cross-spectra likelihood for robust $τ$ constraints from all satellite polarisation data}",
      journal = {arXiv e-prints},
         year = 2026,
        month = mar,
          eid = {arXiv:2603.22454},
        pages = {arXiv:2603.22454},
          doi = {10.48550/arXiv.2603.22454},
archivePrefix = {arXiv},
       eprint = {2603.22454},
 primaryClass = {astro-ph.CO},
       adsurl = {https://ui.adsabs.harvard.edu/abs/2026arXiv260322454G}
}

@ARTICLE{planck2016-l06, 
author = {{\sorthelp{Planck Collaboration 2018F}}{Planck Collaboration VI}},
title = "{\textit{Planck} 2018 results. VI. Cosmological parameters}",
journal = {\aap},
archivePrefix = "arXiv",
eprint = {1807.06209},
year = 2020,
volume = 641,
pages = {A6},
doi = {10.1051/0004-6361/201833910}
}

@ARTICLE{2020A&A...635A..99P,
       author = {{Pagano}, L. and {Delouis}, J. -M. and {Mottet}, S. and {Puget}, J. -L. and {Vibert}, L.},
        title = "{Reionization optical depth determination from Planck HFI data with ten percent accuracy}",
      journal = {\aap},
         year = 2020,
        month = mar,
       volume = {635},
          eid = {A99},
        pages = {A99},
          doi = {10.1051/0004-6361/201936630},
archivePrefix = {arXiv},
       eprint = {1908.09856},
 primaryClass = {astro-ph.CO},
       adsurl = {https://ui.adsabs.harvard.edu/abs/2020A&A...635A..99P}
}

@ARTICLE{2023A&A...675A..12P,
       author = {{Paradiso}, S. and {Colombo}, L.~P.~L. and {Andersen}, K.~J. and {Aurlien}, R. and {Banerji}, R. and {Basyrov}, A. and {Bersanelli}, M. and {Bertocco}, S. and {Brilenkov}, M. and {Carbone}, M. and {Eriksen}, H.~K. and {Eskilt}, J.~R. and {Foss}, M.~K. and {Franceschet}, C. and {Fuskeland}, U. and {Galeotta}, S. and {Galloway}, M. and {Gerakakis}, S. and {Gjerl{\o}w}, E. and {Hensley}, B. and {Herman}, D. and {Iacobellis}, M. and {Ieronymaki}, M. and {Ihle}, H.~T. and {Jewell}, J.~B. and {Karakci}, A. and {Keih{\"a}nen}, E. and {Keskitalo}, R. and {Maggio}, G. and {Maino}, D. and {Maris}, M. and {Partridge}, B. and {Reinecke}, M. and {San}, M. and {Suur-Uski}, A. -S. and {Svalheim}, T.~L. and {Tavagnacco}, D. and {Thommesen}, H. and {Watts}, D.~J. and {Wehus}, I.~K. and {Zacchei}, A.},
        title = "{BEYONDPLANCK. XII. Cosmological parameter constraints with end-to-end error propagation}",
      journal = {\aap},
         year = 2023,
        month = jul,
       volume = {675},
          eid = {A12},
        pages = {A12},
          doi = {10.1051/0004-6361/202244060},
archivePrefix = {arXiv},
       eprint = {2205.10104},
 primaryClass = {astro-ph.CO},
       adsurl = {https://ui.adsabs.harvard.edu/abs/2023A&A...675A..12P}
}

@ARTICLE{planck2020-LVII,
author = {{\sorthelp{Planck Collaboration IntZZG}}{Planck Collaboration Int. LVII}},
title = "{\textit{Planck} intermediate results. LVII. NPIPE: Joint \Planck\ LFI
 and HFI data processing}",
journal = {\aap},
archivePrefix = "arXiv",
eprint = {2007.04997},
year = 2020,
volume = 643,
pages = {A42},
doi = {10.1051/0004-6361/202038073}
}

@ARTICLE{kSZ1,
       author = {{Zeldovich}, Ya. B. and {Sunyaev}, R.~A.},
        title = "{The Interaction of Matter and Radiation in a Hot-Model Universe}",
      journal = {\apss},
         year = 1969,
        month = jul,
       volume = {4},
       number = {3},
        pages = {301-316},
          doi = {10.1007/BF00661821},
       adsurl = {https://ui.adsabs.harvard.edu/abs/1969Ap&SS...4..301Z}
}

@article{kSZ2,
  title={The velocity of clusters of galaxies relative to the microwave background-The possibility of its measurement},
  author={Sunyaev, RA and Zeldovich, Ya B},
  journal={Monthly Notices of the Royal Astronomical Society, vol. 190, Feb. 1980, p. 413-420.},
  volume={190},
  pages={413--420},
  year={1980}
}

@article{Park_Alvarez_Bond_2018,
 title={The Impact of Baryonic Physics on the Kinetic Sunyaev–Zel’dovich Effect}, volume={853}, 
 rights={https://iopscience.iop.org/page/copyright}, 
 ISSN={0004-637X, 1538-4357}, 
 DOI={10.3847/1538-4357/aaa0da}, 
 number={2}, 
 journal={\apj}, 
 publisher={American Astronomical Society}, 
 author={Park, Hyunbae and Alvarez, Marcelo A. and Bond, J. Richard}, 
 year={2018}, 
 month=feb, 
 pages={121}, 
 language={en} }

@ARTICLE{ACTBeringue,
       author = {{Beringue}, Benjamin and {Surrao}, Kristen M. and {Hill}, J. Colin and {Atkins}, Zachary and {Battaglia}, Nicholas and {Bolliet}, Boris and {Calabrese}, Erminia and {Choi}, Steve K. and {Clark}, Susan E. and {Duivenvoorden}, Adriaan J. and {Dunkley}, Jo and {Giardiello}, Serena and {Goldstein}, Samuel and {Hensley}, Brandon S. and {Hlo{\v{z}}ek}, Ren{\'e}e and {Jense}, Hidde T. and {Kramer}, Darby and {La Posta}, Adrien and {Louis}, Thibaut and {Mehta}, Yogesh and {Moodley}, Kavilan and {Naess}, Sigurd and {Partridge}, Bruce and {Qu}, Frank J. and {Ried Guachalla}, Bernardita and {Sehgal}, Neelima and {Sif{\'o}n}, Crist{\'o}bal and {Staggs}, Suzanne T. and {Trac}, Hy and {Van Engelen}, Alexander and {Wollack}, Edward J.},
        title = "{The Atacama Cosmology Telescope: DR6 power spectrum foreground model and validation}",
      journal = {\jcap},
         year = 2025,
        month = oct,
       volume = {2025},
       number = {10},
          eid = {082},
        pages = {082},
          doi = {10.1088/1475-7516/2025/10/082},
archivePrefix = {arXiv},
       eprint = {2506.06274},
 primaryClass = {astro-ph.CO},
       adsurl = {https://ui.adsabs.harvard.edu/abs/2025JCAP...10..082B}
}

@ARTICLE{ACTLouis,
       author = {{Louis}, Thibaut and {La Posta}, Adrien and {Atkins}, Zachary and {Jense}, Hidde T. and {Abril-Cabezas}, Irene and {Addison}, Graeme E. and {Ade}, Peter A.~R. and {Aiola}, Simone and {Alford}, Tommy and {Alonso}, David and {Amiri}, Mandana and {An}, Rui and {Austermann}, Jason E. and {Barbavara}, Eleonora and {Battaglia}, Nicholas and {Battistelli}, Elia Stefano and {Beall}, James A. and {Bean}, Rachel and {Beheshti}, Ali and {Beringue}, Benjamin and {Bhandarkar}, Tanay and {Biermann}, Emily and {Bolliet}, Boris and {Bond}, J. Richard and {Calabrese}, Erminia and {Capalbo}, Valentina and {Carrero}, Felipe and {Chen}, Shi-Fan and {Chesmore}, Grace and {Cho}, Hsiao-mei and {Choi}, Steve K. and {Clark}, Susan E. and {Cothard}, Nicholas F. and {Coughlin}, Kevin and {Coulton}, William and {Crichton}, Devin and {Crowley}, Kevin T. and {Darwish}, Omar and {Devlin}, Mark J. and {Dicker}, Simon and {Duell}, Cody J. and {Duff}, Shannon M. and {Duivenvoorden}, Adriaan J. and {Dunkley}, Jo and {Dunner}, Rolando and {Embil Villagra}, Carmen and {Fankhanel}, Max and {Farren}, Gerrit S. and {Ferraro}, Simone and {Foster}, Allen and {Freundt}, Rodrigo and {Fuzia}, Brittany and {Gallardo}, Patricio A. and {Garrido}, Xavier and {Gerbino}, Martina and {Giardiello}, Serena and {Gill}, Ajay and {Givans}, Jahmour and {Gluscevic}, Vera and {Goldstein}, Samuel and {Golec}, Joseph E. and {Gong}, Yulin and {Guan}, Yilun and {Halpern}, Mark and {Harrison}, Ian and {Hasselfield}, Matthew and {Healy}, Erin and {Henderson}, Shawn and {Hensley}, Brandon and {Herv{\'\i}as-Caimapo}, Carlos and {Hill}, J. Colin and {Hilton}, Gene C. and {Hilton}, Matt and {Hincks}, Adam D. and {Hlo{\v{z}}ek}, Ren{\'e}e and {Ho}, Shuay-Pwu Patty and {Hood}, John and {Hornecker}, Erika and {Huber}, Zachary B. and {Hubmayr}, Johannes and {Huffenberger}, Kevin M. and {Hughes}, John P. and {Ikape}, Margaret and {Irwin}, Kent and {Isopi}, Giovanni and {Joshi}, Neha and {Keller}, Ben and {Kim}, Joshua and {Knowles}, Kenda and {Koopman}, Brian J. and {Kosowsky}, Arthur and {Kramer}, Darby and {Kusiak}, Aleksandra and {Lagu{\"e}}, Alex and {Lakey}, Victoria and {Lee}, Eunseong and {Li}, Yaqiong and {Li}, Zack and {Limon}, Michele and {Lokken}, Martine and {Lungu}, Marius and {MacCrann}, Niall and {MacInnis}, Amanda and {Madhavacheril}, Mathew S. and {Maldonado}, Diego and {Maldonado}, Felipe and {Mallaby-Kay}, Maya and {Marques}, Gabriela A. and {van Marrewijk}, Joshiwa and {McCarthy}, Fiona and {McMahon}, Jeff and {Mehta}, Yogesh and {Menanteau}, Felipe and {Moodley}, Kavilan and {Morris}, Thomas W. and {Mroczkowski}, Tony and {Naess}, Sigurd and {Namikawa}, Toshiya and {Nati}, Federico and {Nerval}, Simran K. and {Newburgh}, Laura and {Nicola}, Andrina and {Niemack}, Michael D. and {Nolta}, Michael R. and {Orlowski-Scherer}, John and {Pagano}, Luca and {Page}, Lyman A. and {Pandey}, Shivam and {Partridge}, Bruce and {Perez Sarmiento}, Karen and {Prince}, Heather and {Puddu}, Roberto and {Qu}, Frank J. and {Ragavan}, Damien C. and {Ried Guachalla}, Bernardita and {Rogers}, Keir K. and {Rojas}, Felipe and {Sakuma}, Tai and {Schaan}, Emmanuel and {Schmitt}, Benjamin L. and {Sehgal}, Neelima and {Shaikh}, Shabbir and {Sherwin}, Blake D. and {Sierra}, Carlos and {Sievers}, Jon and {Sif{\'o}n}, Crist{\'o}bal and {Simon}, Sara and {Sonka}, Rita and {Spergel}, David N. and {Staggs}, Suzanne T. and {Storer}, Emilie and {Surrao}, Kristen and {Switzer}, Eric R. and {Tampier}, Niklas and {Thornton}, Robert and {Trac}, Hy and {Tucker}, Carole and {Ullom}, Joel and {Vale}, Leila R. and {Van Engelen}, Alexander and {Van Lanen}, Jeff and {Vargas}, Cristian and {Vavagiakis}, Eve M. and {Wagoner}, Kasey and {Wang}, Yuhan and {Wenzl}, Lukas and {Wollack}, Edward J. and {Zheng}, Kaiwen and {The Atacama Cosmology Telescope collaboration}},
        title = "{The Atacama Cosmology Telescope: DR6 power spectra, likelihoods and {\ensuremath{\Lambda}}CDM parameters}",
      journal = {\jcap},
         year = 2025,
        month = nov,
       volume = {2025},
       number = {11},
          eid = {062},
        pages = {062},
          doi = {10.1088/1475-7516/2025/11/062},
archivePrefix = {arXiv},
       eprint = {2503.14452},
 primaryClass = {astro-ph.CO},
       adsurl = {https://ui.adsabs.harvard.edu/abs/2025JCAP...11..062L}
}

@ARTICLE{Agora,
       author = {{Omori}, Yuuki},
        title = "{AGORA: Multicomponent simulation for cross-survey science}",
      journal = {\mnras},
         year = 2024,
        month = jun,
       volume = {530},
       number = {4},
        pages = {5030-5068},
          doi = {10.1093/mnras/stae1031},
archivePrefix = {arXiv},
       eprint = {2212.07420},
 primaryClass = {astro-ph.CO},
       adsurl = {https://ui.adsabs.harvard.edu/abs/2024MNRAS.530.5030O}
}

@article{Douspis_2022_I,
   title={Retrieving cosmological information from small-scale CMB foregrounds: I. The thermal Sunyaev Zel’dovich effect},
   volume={659},
   ISSN={1432-0746},
   url={http://dx.doi.org/10.1051/0004-6361/202142004},
   DOI={10.1051/0004-6361/202142004},
   journal={Astronomy \& Astrophysics},
   publisher={EDP Sciences},
   author={Douspis, Marian and Salvati, Laura and Gorce, Adélie and Aghanim, Nabila},
   year={2022},
   month=mar, pages={A99} }

@ARTICLE{GorceDouspis_2022,
       author = {{Gorce}, Ad{\'e}lie and {Douspis}, Marian and {Salvati}, Laura},
        title = "{Retrieving cosmological information from small-scale CMB foregrounds. II. The kinetic Sunyaev Zel'dovich effect}",
      journal = {\aap},
         year = 2022,
        month = jun,
       volume = {662},
          eid = {A122},
        pages = {A122},
          doi = {10.1051/0004-6361/202243351},
archivePrefix = {arXiv},
       eprint = {2202.08698},
 primaryClass = {astro-ph.CO},
       adsurl = {https://ui.adsabs.harvard.edu/abs/2022A&A...662A.122G}
}

@ARTICLE{Zahn2012_reio,
       author = {{Zahn}, O. and {Reichardt}, C.~L. and {Shaw}, L. and {Lidz}, A. and {Aird}, K.~A. and {Benson}, B.~A. and {Bleem}, L.~E. and {Carlstrom}, J.~E. and {Chang}, C.~L. and {Cho}, H.~M. and {Crawford}, T.~M. and {Crites}, A.~T. and {de Haan}, T. and {Dobbs}, M.~A. and {Dor{\'e}}, O. and {Dudley}, J. and {George}, E.~M. and {Halverson}, N.~W. and {Holder}, G.~P. and {Holzapfel}, W.~L. and {Hoover}, S. and {Hou}, Z. and {Hrubes}, J.~D. and {Joy}, M. and {Keisler}, R. and {Knox}, L. and {Lee}, A.~T. and {Leitch}, E.~M. and {Lueker}, M. and {Luong-Van}, D. and {McMahon}, J.~J. and {Mehl}, J. and {Meyer}, S.~S. and {Millea}, M. and {Mohr}, J.~J. and {Montroy}, T.~E. and {Natoli}, T. and {Padin}, S. and {Plagge}, T. and {Pryke}, C. and {Ruhl}, J.~E. and {Schaffer}, K.~K. and {Shirokoff}, E. and {Spieler}, H.~G. and {Staniszewski}, Z. and {Stark}, A.~A. and {Story}, K. and {van Engelen}, A. and {Vanderlinde}, K. and {Vieira}, J.~D. and {Williamson}, R.},
        title = "{Cosmic Microwave Background Constraints on the Duration and Timing of Reionization from the South Pole Telescope}",
      journal = {\apj},
         year = 2012,
        month = sep,
       volume = {756},
       number = {1},
          eid = {65},
        pages = {65},
          doi = {10.1088/0004-637X/756/1/65},
archivePrefix = {arXiv},
       eprint = {1111.6386},
 primaryClass = {astro-ph.CO},
       adsurl = {https://ui.adsabs.harvard.edu/abs/2012ApJ...756...65Z}
}

@article{kSZmeasurement2,
  title={An improved measurement of the secondary cosmic microwave background anisotropies from the SPT-SZ+ SPTpol surveys},
  author={Reichardt, CL and Patil, S and Ade, PAR and Anderson, AJ and Austermann, JE and Avva, JS and Baxter, E and Beall, JA and Bender, AN and Benson, BA and others},
  journal={The Astrophysical Journal},
  volume={908},
  number={2},
  pages={199},
  year={2021},
  publisher={IOP Publishing}
}

@ARTICLE{Battaglia,
       author = {{Battaglia}, N. and {Natarajan}, A. and {Trac}, H. and {Cen}, R. and {Loeb}, A.},
        title = "{Reionization on Large Scales. III. Predictions for Low-l Cosmic Microwave Background Polarization and High-l Kinetic Sunyaev-Zel'dovich Observables}",
      journal = {\apj},
         year = 2013,
        month = oct,
       volume = {776},
       number = {2},
          eid = {83},
        pages = {83},
          doi = {10.1088/0004-637X/776/2/83},
archivePrefix = {arXiv},
       eprint = {1211.2832},
 primaryClass = {astro-ph.CO},
       adsurl = {https://ui.adsabs.harvard.edu/abs/2013ApJ...776...83B}
}

@Article{matplotlib,
  Author    = {Hunter, J. D.},
  Title     = {Matplotlib: A 2D graphics environment},
  Journal   = {Computing in Science \& Engineering},
  Volume    = {9},
  Number    = {3},
  Pages     = {90--95},
  publisher = {IEEE COMPUTER SOC},
  doi       = {10.1109/MCSE.2007.55},
  year      = 2007
}

@article{Lewis:1999bs,
      author         = "Lewis, Antony and Challinor, Anthony and Lasenby,
                        Anthony",
      title          = "{Efficient computation of CMB anisotropies in closed FRW
                        models}",
      journal        = "\apj",
      volume         = "538",
      year           = "2000",
      pages          = "473-476",
      doi            = "10.1086/309179",
      eprint         = "astro-ph/9911177",
      archivePrefix  = "arXiv",
      primaryClass   = "astro-ph",
      SLACcitation   = "%%CITATION = ASTRO-PH/9911177;%%"
}

@ARTICLE{Whitler,
       author = {{Whitler}, Lily and {Stark}, Daniel P. and {Topping}, Michael W. and {Robertson}, Brant and {Rieke}, Marcia and {Hainline}, Kevin N. and {Endsley}, Ryan and {Chen}, Zuyi and {Baker}, William M. and {Bhatawdekar}, Rachana and {Bunker}, Andrew J. and {Carniani}, Stefano and {Charlot}, St{\'e}phane and {Chevallard}, Jacopo and {Curtis-Lake}, Emma and {Egami}, Eiichi and {Eisenstein}, Daniel J. and {Helton}, Jakob M. and {Ji}, Zhiyuan and {Johnson}, Benjamin D. and {P{\'e}rez-Gonz{\'a}lez}, Pablo G. and {Rinaldi}, Pierluigi and {Tacchella}, Sandro and {Williams}, Christina C. and {Willmer}, Christopher N.~A. and {Willott}, Chris and {Witstok}, Joris},
        title = "{The z {\ensuremath{\gtrsim}} 9 Galaxy UV Luminosity Function from the JWST Advanced Deep Extragalactic Survey: Insights into Early Galaxy Evolution and Reionization}",
      journal = {\apj},
         year = 2025,
        month = oct,
       volume = {992},
       number = {1},
          eid = {63},
        pages = {63},
          doi = {10.3847/1538-4357/adfddc},
archivePrefix = {arXiv},
       eprint = {2501.00984},
 primaryClass = {astro-ph.GA},
       adsurl = {https://ui.adsabs.harvard.edu/abs/2025ApJ...992...63W}
}

\appendix

\onecolumn

\section{Properties of reionisation histories}

\begin{table*}[hb!]
\caption{Summary characteristics of each best {\it Planck} PR4 reionisation histories. }                 
\label{table:A1}    
\centering     
\begin{tabular}{ccccccccc}
\hline\hline
Model & Type & $z_{start}$ & $\Delta_{10-90}$ & $\tau_{z>10}$ (\%) & $D_{2000}^{kSZ}$ ($\mu K^2$) & $D_{2000}^{kSZ} (z>10)$ (\%) & $\chi^2$ (free CIB-SZ) & $\chi^2$ (free CIB) \\
\hline
3-knots & early & 16.82 & 11.52 & 29.95 & 5.33 & 25.64 & 10.83 & 32.86 \\
10-bins & early & 17.08 & 12.14 & 38.96 & 4.55 & 26.37 & 5.20 & 21.14 \\
2-steps & early & 15.01 & 9.29 & 33.17 & 4.69 & 23.88 & 6.06 & 23.06 \\
3-steps & early & 17.48 & 11.73 & 37.71 & 4.16 & 24.81 & 3.30 & 16.26 \\ \hline
2-pow & late & 9.00 & 1.47 & 0.29 & 2.20 & 0.22 & 2.66 & 2.48 \\
3-flex & late & 8.19 & 0.27 & 0.11 & 1.90 & 0.08 & 3.83 & 1.77 \\
2-flex & late & 8.21 & 0.32 & 0.11 & 1.92 & 0.08 & 3.75 & 1.80 \\
tanh & late & 8.63 & 0.97 & 0.11 & 2.07 & 0.08 & 3.19 & 2.07 \\
2-knots & late & 8.29 & 0.40 & 0.11 & 1.94 & 0.08 & 3.66 & 1.84 \\
tanhdz & late & 8.16 & 0.25 & 0.11 & 1.91 & 0.08 & 3.79 & 1.79 \\
3-pow & late & 8.19 & 0.22 & 0.11 & 1.91 & 0.08 & 3.75 & 1.81 \\
\hline    
\end{tabular}
\tablefoot{$z_{10}^{first}$ is defined as the redshift when the reionisation history reaches
10\% for the first time while  $\Delta_{10-90}$ is defined as the interval between the first time $x_e(z)$ reaches 10\% and the last time it reaches 90\%. $\tau_{z>10}$ is defined as the percentage contribution of redshifts above 10 in the computation of the total reionisation optical depth.  $D_{2000}^{kSZ}$ is the amplitude of the kSZ angular power spectra at $\ell=2000$ in $\mu K^2$. $D_{2000}^{kSZ} (z>10)$ shows the contribution in percentage of redshifts above 10 in the amplitude of the kSZ angular power spectrum at $\ell=2000$ . $\chi^2$ (free CIB-SZ) and (free CIB) are the $\chi^2$ value for each model computed using the SPT-3G free CIB-SZ (green dots) and SPT-3G free CIB (green crosses) points in Fig.~\ref{Fig:fig2}  \citep{Chaubal26}.}
\end{table*}

Table \ref{table:A1} summarises the principal characteristics of each model helping to defined the 2 classes: `late' and `early'. The last two columns show the $\chi^2$ values simply computed using the particular cases of `SPT-3G free CIB-SZ' data points (the ones with less assumptions on foregrounds) and `SPT-3G free CIB' data points shown in Fig.~\ref{Fig:fig2}, from \cite{Chaubal26}.

\FloatBarrier 
\twocolumn

\end{document}